# Artificial Micrometeorites

*S. N. Dolya*

*Joint Institute for Nuclear Research, Joliot Curie str. 6, Dubna, Russia, 141980*

**Abstract**

An iron ball, a beryllium sphere and a tungsten tube segment with diameter d = 20 µ, are electrically charged while proton beam irradiating. These bodies are accelerated by the running pulse field in a spiral waveguide up to velocity: $V_m$ = 30 km / s. The accelerator, generating micrometeorites is placed at satellites on the Earth orbit. This article considers processes of penetration of micrometeorites into the Earth atmosphere. It is shown that micrometeorites evaporate at the height of 100 - 150 km from the surface of the Earth. A micrometeorite which is a segment of the beryllium tube equipped with a graphite cone in the head part is the very meteorite to reach the Earth surface without being broken.

**Introduction**

Artificial micrometeorites, which are electrically charged and accelerated by electrodynamics method, may be applied to study fluctuations in the atmospheric density and distribution of the magnetic and electric fields in it. We will consider micrometeorites, consisting of various elements and having a different shape, but they will all have the same diameter $d_m$ = 20 µ and the same velocity: $V_m$ = 30 km / s.

We assume that the accelerator generating micrometeorites is located on a satellite rotating over the near-by orbit of the Earth.

**1. Acceleration of iron balls**

We consider how to accelerate the iron ball having diameter $d_{Fe}$ = 20 µ up to velocity $V_m$ = 30 km / s, or if the velocity is expressed in the units of the velocity of light, $\beta_{fin}$ = $10^{-4}$, where $\beta$ = V / c, c = 3 * $10^5$ km / s, the velocity of light in vacuum.

To speed up this ball by the field of the electromagnetic wave, it must be electrically charged.

It is possible to charge the iron ball electrically by electrons bombarding it, which should stay on the ball thus transferring the electric charge to it. The field emission will limit the amount of electrons on the ball in the process. Starting from the certain field strength due to the field emission the set of electrons will leak out from the ball no matter how many electrons and new electrons we would plant, they will all leave the ball.

The value of the surface threshold for the current density of iron is: e$\varphi$ = 4.3 eV [1], page 444, and the electric field strength $E_{surf}$ = 30 MV / cm in this case is j = $10^{-1}$ A/cm$^2$ [1], page 461. If you cover the iron ball with platinum and passivate it with oxygen, the surface threshold will increase up to e$\varphi$ = 6.5 eV [1], page 445, and the leakage current density (for values $E_{surf}$ = 30 MV /cm will reduce to: j = $10^{-9}$ A/cm$^2$ [1], page 461.

However, with increasing field strength up to 40 MV / cm the leakage current density will increase to j = $10^{-4}$ A/cm$^2$, and at higher growth of the surface field strength up to value $E_{surf}$ = 50 MV / cm, the current density will increase to the value of j = $10^{-1}$ A/cm$^2$ [1], page 461. So that, in achieving this field strength, all new electrons planted on the ball will leak out due to the field emission.

It is evident that it is impossible to achieve the high field strength by placing a large electric charge on the ball, when it is irradiated by the electron beam. This is due to the fact that the electron is a light particle and it easily leaves the ball which has a large surface electric field.

The solution is as follows: it is necessary to irradiate a ball with heavy particles such as protons. We will proceed from the surface electric field $E_{surf}$ = 300 MV / cm. It should be borne in mind that the proton is a heavy particle and while irradiating the ball with protons the ball will be transferred a large transverse (relative to the direction of acceleration) pulse. Therefore, irradiation of the ball must be performed by two opposite beams of protons to exclude the transverse pulse.

*1. 1. The number of protons placed on the ball*

We find the number of protons, placed on the ball, on the Coulomb law:

$$E_{surf} = N_p \, e/r_{Fe}^2, \qquad (1)$$

where $E_{surf}$ = 300 MV / cm is the surface field strength on the ball,
e = 4.8 * $10^{-10}$ - elementary charge, expressed in CGS units,
$r_{Fe}$ = 10 µ - ball radius, $N_p$ - the number of protons, placed on the ball.

Substituting numbers into the formula (1) we find that it is required to place $N_p = 2 * 10^9$ protons to obtain the surface field strength $E_{surf} = 300$ MV/cm.

For the protons would overcome the electrostatic repulsion of the protons previously placed on the ball, the proton energy must be higher than:

$$W_p > E_{surf} * r_{Fe} = 300 \text{ keV}. \qquad (2)$$

The running length of the protons with the energy of 300 keV in iron can be estimated as $R_{Fe} < 1$ μ [1], page 953, which is less than the radius of the ball. If the running length is greater than the radius of the ball, the proton energy should be increased gradually having been slowed down by the Coulomb field of the ball and the protons would remain on the ball.

*1. 2. Number of nucleons in the ball*

In the accelerator technology an important parameter is the ratio of charge to the mass of the accelerated particle where the charge is expressed in units of the elementary charge (electron, proton), and the mass is expressed in the units of the mass of the nucleon.

We find the mass of the iron ball. The volume of the ball is: $V_{Fe} = (4/3) \pi r^3_{Fe}$, $V \approx 4 * 10^{-9}$ cm$^3$, the density of iron is assumed to be: $\rho_{Fe} = 8$ g/cm$^3$, so that the mass of the ball is $m_{Fe} \approx 3 * 10^{-8}$ g. The number of nucleons in the ball is found from Avogadro's law:

$$6*10^{23} \text{ ------------ } 56 \text{ g}$$
$$\text{x ------------} 3*10^{-8} \text{ g}, \qquad (3)$$

where: 56 g - mass of one gram mole of iron. From (3) we find that
$x = 6 * 10^{23} * 3 * 10^{-8}/56$ atoms or $A = 6 * 10^{23} * 3 * 10^{-8} \approx 2 * 10^{16}$ nucleons.

The ratio of the charge to the mass (Z / A) for the ball is as follows:

$$eN_p/A = Z/A = 2*10^9/2*10^{16} = 10^{-7}. \qquad (4)$$

*1. 3. Electrostatic acceleration of the balls*

Usually linear accelerators of heavy charged particles are constructed as follows. First, preliminary electrostatic acceleration is performed at a certain speed, till which the electromagnetic wave can be slowed down and further the final acceleration of the traveling wave is performed. In our case, since the ratio of the charge to the mass of the ball is extremely small ($10^{-7}$) it is needed to uses a structure where the ultra slow electromagnetic waves can propagate. Recall that the singly charged ion of Uranium - 238 has a ratio of the charge to the mass $1/238 = 4.2 * 10^{-3}$ that is much bigger than in our case. Let us select the voltage for the platform on where the store with the balls is located, to be equal to $U_{el.\ st.} = 100$ kV.

The balls from the store should be produced and mechanically accelerated till the velocity of about one meter per second. Thereafter, during the motion they must be irradiated by two opposite proton beams so that the number of protons placed on the ball should be equal to $N_p = 2 * 10^9$. After that, the balls must be accelerated by electrostatic potential difference $U_{el.\ st.} = 100$ kV.

Let the field strength of the iron ball be $E_{el.\ st.} = 10$ kV / cm. The equation of motion of the ball with a specific (per nucleon) charge $Z / A = 10^{-7}$, can be written as follows:
$$dV/dt = (Z/A)eE_{el.\ st.}/M, \qquad (5)$$

where M is the mass of the nucleon. Assuming the initial velocity of the ball is equal to zero, we obtain an expression for the dimensionless velocity $\beta = V / c$, where $c = 3 \times 10^5$ km / s - the velocity of light in vacuum. We write the dependence of the dimensionless velocity of time:

$$\beta = \{(Z/A)*eE_{el.\ st.}/Mc^2\}*ct. \qquad (6)$$

The final (electrostatic acceleration) velocity can be found from the following relation:

$$\beta^2 = 2(Z/A)e\ U_{el.\ st}/Mc^2, \qquad (7)$$

where we find: $\beta^2 = 2*10^{-11}$, $\beta = 4.5*10^{-6}$.

*1. 4. Electromagnetic acceleration of iron balls*

We choose a spiral waveguide as the slow-wave structure, which can propagate ultra slow electromagnetic waves [2].

*1. 4. 1. The required delay of the electromagnetic wave*

To speed up charged particles by using the electrodynamics way, it is necessary that the initial velocity of the particles and waves coincide approximately: $\beta = \beta_{ph}$, and it is required to increase the phase velocity of the wave while particle acceleration.

The slowdown of the waves in the tight helix is approximately equal to the ratio of the perimeter to the pitch of the helix. The dispersion equation relating the phase velocity with the size of the spiral $\beta_{ph}$ in the tight helix looks like this:

$$\beta_{ph} = \text{tg}\Psi, \qquad (8)$$

where $\beta_{ph} = V_{ph}/c$, $\text{tg}\Psi = h/2\pi r_0$ - tangent of the winding angle, h - the winding pitch of the helix, $r_0$ - the radius of the spiral.

It is also known that in the medium where the permittivity and permeability have significant values, the phase velocity of the electromagnetic wave is less than the velocity of light in vacuum and equal to the following:

$$\beta_{ph} = 1/(\varepsilon\mu)^{1/2}. \qquad (9)$$

We can expect that if you put the spiral into this medium, the general slowdown of the electrostatic wave will increase and the electromagnetic wave phase velocity will be equal to:

$$\beta_{ph} = \text{tg}\Psi/(\varepsilon\mu)^{1/2}, \qquad (10)$$

where factor $\text{tg}\Psi$ determines the slowing properties of the structure and $(\varepsilon\mu)^{1/2}$ - shows what properties of slowdown the medium possesses. It is evident that considerable quantities $(\varepsilon\mu)^{1/2}$, according to formula (10) may be prepared by slowing by the order of $10^5$.

*1. 4. 2. Exact solutions*

Now we consider the properties of helical waveguide, fully immersed to the medium with permittivity ε and permeability μ. From Maxwell's equations for the internal components of the field area we find the following field components for the inner area of the spiral:

$$E_{z1} = E_0 I_0(k_1 r)$$
$$E_{r1} = i(k_3/k_1) E_0 I_1(k_1 r)$$
$$H_{\varphi 1} = i\varepsilon(k/k_1) E_0 I_1(k_1 r)$$
$$H_{z1} = -i(k_1/\mu k) tg\Psi I_0(k_1 r_0) E_0 I_0(k_1 r)/I_1(k_1 r_0)$$
$$E_{\varphi 1} = -tg\Psi I_0(k_1 r_0) E_0 I_1(k_1 r)/I_1(k_1 r_0)$$
$$H_{r1} = (k_3/\mu k) tg\Psi I_0(k_1 r_0) E_0 I_1(k_1 r)/I_1(k_1 r_0). \quad (11)$$

The field components for the helix external area can be written in the form [2]:

$$E_{z2} = I_0(k_1 r_0) E_0 K_0(k_1 r)/K_0(k_1 r_0)$$
$$E_{r2} = -i(k_3/k_1) I_0(k_1 r_0) E_0 K_1(k_1 r)/K_0(k_1 r_0)$$
$$H_{\varphi 2} = -i(k/k_1) I_0(k_1 r_0) E_0 K_1(k_1 r)/K_0(k_1 r_0)$$
$$H_{z2} = i(k_1/k) tg\Psi I_0(k_1 r_0) E_0 K_0(k_1 r)/K_1(k_1 r_0)$$
$$E_{\varphi 2} = -tg\Psi I_0(k_1 r_0) E_0 K_1(k_1 r)/K_1(k_1 r_0)$$
$$H_{r2} = (k_3/\mu k) tg\Psi I_0(k_1 r_0) E_0 K_1(k_1 r)/K_1(k_1 r_0), \quad (12)$$

where the omitted factor is $e^{i(\omega t - k_3 z)}$.

The dispersion equation, the relationship between the phase velocity of the wave, the structure parameters and the wave frequency for the helix, fully immersed in Ferro dielectric medium is as follows:

$$ctg^2\Psi = k_1^2/k^2 \{I_0(k_1 r_0) K_0(k_2 r_0)/I_1(k_1 r_0) K_1(k_2 r_0)\}, \quad (13)$$

where: $k_1 = k(1/\beta_{ph}^2 - 1)^{1/2}$, $k_2 = k(1/\beta_{ph}^2 - \varepsilon\mu)^{1/2}$, $k = \omega/[c(\varepsilon\mu)^{1/2}]$. For the large slow down, the dispersion equation is simplified and looks like this:

$$\beta_{ph} = tg\Psi /(\varepsilon\mu)^{1/2}. \quad (14)$$

This formula, as well as for the helix located in vacuum has a simple physical meaning: the ratio of the phase velocity to the velocity of the wave in the medium is the ratio of the lengths to the wave running along the helix and its axis. After simple calculations we obtain the relation between the power flux in the spiral and $E_0$ field tension on its axis in the case of the spiral emerged into medium c with permittivity ε [2]:

$$P = (c/8)\, E_0^2\, r_0^2\, [\,kk_3/k_1^2\,]\, \varepsilon\{(1+I_0K_1/I_1K_0)(I_1^2-I_0I_2) +$$

$$+ (I_0/K_0)^2(1+I_1K_0/I_0K_1)(K_0K_2-K_1^2)\}. \qquad (15)$$

This relationship between the flux and the electric field strength on the axis coincides with the expression for the spiral in vacuum except for factor ε which determines the dielectric constant of the medium.

*1. 4. 3. Partial filling of the medium*

If the insulator is only outside the helix region and inside the helix is free of the filling, formula (15), relating the power flux with the electric field strength on the axis, is somewhat different [2]:

$$P = (c/8)\, E_0^2\, r_0^2[\,kk_3/k_1^2\,]\{(1+I_0K_1/I_1K_0)(I_1^2-I_0I_2) +$$

$$+ \varepsilon\, (I_0/K_0)^2(1+I_1K_0/I_0K_1)(K_0K_2-K_1^2)\}. \qquad (16)$$

Only the second term in the curly brackets, the corresponding power flux propagating beyond the spiral is multiplied by ε. The factor ε before the bracket does not contain it. We call this case - partial filling of the spiral with the medium.

The dispersion equation relating the phase velocity of the wave with frequency is as follows if the moderating medium is beyond the helix [2]:

$$\operatorname{ctg}^2\Psi = (k_1k_2/k^2)\varepsilon\mu\{I_0(k_1r_0)K_0(k_2r_0)/I_1(k_1r_0)K_1(k_2r_0)\}F_0, \qquad (17)$$

where:

$$F_0 = \varepsilon\{1+(k_1/k_2)\mu\, I_0K_1/I_1K_0\}*[1+(k_1/k_2)\varepsilon\, I_0K_1/I_1K_0]^{-1}. \qquad (18)$$

Arguments of the functions $I_{0,\,1}$ are: $(k_1r_0)$, functions $K_{0,\,1}$ are: $(k_2r_0)$.
The most interesting case for us is tight winding, thus, the dispersion equation for this case is simplified and looks as follows:

$$\beta_{ph} = tg\Psi \, F_0^{1/2}/(\varepsilon\mu)^{1/2}. \tag{19}$$

It is easy to see that in the case of $\varepsilon, \mu = 1$, the dispersion equation becomes the vacuum case: $\beta_{ph} = tg\Psi$. In the general case the $F_0^{1/2} > 1$ and retarding properties in the spiral filled Ferro dielectrics outside are worse than in the case with full immersion.

In the important case of $\varepsilon, \mu \gg 1$ the expression for $F_0$ is simplified:

$$F_0 = \mu, \tag{20}$$

and the dispersion equation becomes:

$$\beta_{ph} = tg\Psi/\varepsilon^{1/2}. \tag{21}$$

In the most interesting case is when large dielectric $\varepsilon$, $\varepsilon \gg 1$ is outside the helix and the medium is not ferromagnetic $\mu = 1$, the expression is simplified for $F_0 = 2$, and the dispersion equation looks like this:

$$\beta_{ph} = \sqrt{2}\, tg\Psi/\varepsilon^{1/2}. \tag{22}$$

This very case providing the large slowdown and high electric field $E_0$ on the axis will be considered below in detail.

Maintaining synchronism between the wave velocity and particle velocity can be achieved by increasing the helix pitch. For a spiral wound on the cylindrical surface it would lead to thin winding and that is why the electric field strength (for a fixed power generator) will reduce. There is another variant of making the spiral waveguide. It is possible to wind the spiral onto the narrowing cone. In fact in this case there is an increase of the phase velocity achieved not by increasing the helical pitch h, but by reducing radius $r_0$. At the same factor ( $kk_3/k^2_1$ ) $r^2_0$, equal to $\beta r^2_0$, remains approximately constant.

*1. 4. 4. Acceleration length*

Finite velocity $V_m = 30$ km / s corresponds to the velocity $\beta_{fin}$, expressed in units of the velocity of light, $c = 3 * 10^5$ km / s, which is equal to: $\beta_{fin} = 10^{-4}$. The energy per nucleon in this case is: $W_{fin} = Mc^2 \beta^2_{fin} / 2 = 5$ eV / nucleon, where $Mc^2 = 1$ GeV, the nucleon mass, expressed in energy units. When acceleration rate $eE_0\sin\varphi_s = 250$ keV / cm, where $E_0$ – the peak value of the field strength, $E_0 = 300$ kV / cm, $\varphi_s = 60^0$ synchronous phase, $\sin\varphi_s = 0.87$, this energy can be achieved over the length: $L_{acc\ Fe} = W_{fin} / [(Z / A)\ eE_0\sin\varphi_s] = 2$ m.

*1. 4. 5. Power consumption*

The relationship between the power flux and the electric field strength for a spiral when the space between the spiral and the screen filled with a dielectric and the screen is given by (16).

The argument of the modified Bessel functions in curly brackets in (16) is the value of $x = 2\pi r_0/\lambda_{slow}$, where $r_0$ - radius of the helix, $\lambda_{slow} = \beta\lambda_0$, $\beta$ - phase velocity in a spiral, $\lambda_0 = s/f_0$ - wavelength in free space, $f_0 = \omega/2\pi$ - wave frequency, $\omega$ - angular frequency of the wave.

From formula (16) we see that in order to have the maximum field strength $E_0$ for the fixed power flux P and at fixed $r_0$ and $\beta$, we must have the smallest value of the curly bracket which is simply a numerical factor in the formula (16).

During acceleration there is the rate change $\beta$ of the ball and it leads to the fact that for the given power the electric field strength decreases. To maintain a uniform rate of acceleration within the same section (and as will be seen later, the whole accelerator should be divided into separate sections), it is necessary to reduce the radius of the spiral winding along the length of the section [2]. To reach this, it is needed to wind the spiral onto the narrowing cone so that the product $\beta r^2_0$ remains approximately constant over the length of the section.

Then, at a reasonable length of the section, parameter $x = 2\pi r_0/\beta\lambda_0$ rapidly decreases This is due to the fact that not only the velocity of the ball $\beta$ increases but also the radius of the spiral winding $r_0$ reduces. Table 1 shows the values of the first (I) and second (II) terms in the curly brackets of equation (16).

Table 1.

| x | I | II |
|---|---|---|
| 0.1 | 0.1 | 66.8 |
| 0.2 | 0.14 | 22 |
| 0.3 | 0.18 | 12.14 |
| 0.4 | 0.226 | 8.286 |
| 0.5 | 0.273 | 6.365 |
| 0.6 | 0.326 | 5.277 |
| 0.7 | 0.386 | 4.618 |
| 0.8 | 0.454 | 4.208 |
| 0.9 | 0.532 | 3.958 |
| 1 | 0.620 | 3.819 |
| 1.1 | 0.721 | 3.763 |
| 1.2 | 0.836 | 3.774 |
| 1.3 | 0.968 | 3.844 |
| 1.4 | 1.119 | 3.96 |
| 1.5 | 1.29 | 4.142 |
| 1.6 | 1.494 | 4.369 |
| 1.7 | 1.724 | 4.650 |
| 1.8 | 1.989 | 4.989 |
| 1.9 | 2.295 | 5.393 |
| 2 | 2.69 | 5.867 |
| 2.5 | 5.441 | 9.68 |
| 3 | 11.336 | 17.601 |

Recall that the second term in the curly brackets corresponds to the power flux propagating between the spiral and the screen. This volume should be filled with a dielectric having a large dielectric constant ε, so that the second term in the curly brackets would be much larger than the first one. Acceleration can begin within one section at a value of x = 2, and complete at x = 0.5.

The maximum field strength will then be obtained for a value of x = 1, that is in the middle of the section. The field strength at the beginning and end of section will be less than [2]. Everything discussed above, applied to the coil without an external screen. The exact formulas that take into account the effect of the screen, obtained in [3]. Clear, however, that if the screen is far enough away from the spiral $R_{screen} > 3\ r_0$, then its effect on the propagation of radio waves in the spiral slightly.

We choose the parameter x to x = 2 and calculate the power required to create the field $E_0 = 300$ kV / cm for the beginning of the acceleration, i.e. $\beta_{in} = 4.5 * 10^{-6}$. We choose to determine the initial spiral radius $r_0 = 1$ cm. Value x = 2 means that we chose $2\pi r_0/\beta_{in}\lambda_0 = 2$, $\lambda_{slow} = \pi r_0 = 3.14$ cm, then we have chosen to start accelerating the vacuum wavelength $\lambda_0 = \pi r_0/\beta_{in} = 7 * 10^5$ cm. This wavelength corresponds to a frequency $f_0 = c / \lambda_0 = 43$ kHz.

As we have already noted, besides the purely geometric spiral wave slowdown the wave should be additionally slowed down further. To do this, the space between the spiral and the outer shield must be filled with dielectric medium having relative permittivity $\varepsilon$. For barium titanate, near the Curie point, the achievable values of $\varepsilon = 8 * 10^3$, [4], page 557. We take value $\varepsilon$ slightly smaller, namely: $\varepsilon = 1.28 * 10^3$.

Substitute the numbers into the formula (16).

$P (W) = 3*10^{10}*3*10^5*3*10^5*4.5*10^{-6}*1.28*10^3*5.87/8*3*10^2*3*10^2*10^7 =$
$= 12.6 *10^6$ W.

There is a large number of high-frequency generators having this electric power, obtained in the previous expression. However, if to accelerate a single bulb, it can be accelerated by the field travelling pulse along the spiral waveguide. Duration of the base of the sinusoidal pulse corresponding to frequency $f_0$, is: $\tau_{pulse} = 1/2f_0 \approx 10$ μs. The pulse amplitude can be found from the following relation:

$$U_{pulse} = E_0 * \lambda_{slow}/2\pi = 314 \text{ kV}. \qquad (23)$$

*1. 4. 6. The spiral pitch*

Since we have chosen the spiral radius equal to: $r_0 = 1$ cm, then it is required to take a very small spiral winding pitch to get the slowdown equal to $\beta_{ph\ in} = 4.5 * 10^{-6}$ in this spiral, where $\beta_{ph\ in} = V_{in} / c$ - initial phase velocity, expressed in terms of the velocity of light coinciding with the initial velocity of the ball.

According to formula (22) for the start of the spiral where the velocity of the balls is $\beta_{ph\ in} = 4.5 * 10^{-6}$, $r_0 = 1$ cm, $\varepsilon = 1.28 * 10^3$, replacing tg $\Psi \approx h/2\pi r_0$, where h - the helix winding pitch, we find,

$$4.5*10^{-6} = \sqrt{2}*h/(2\pi r_0 * \varepsilon^{1/2}).$$

From this expression we define that the helix winding pitch should be equal to:

$$h = 7 \text{ μ}. \qquad (24)$$

The amplitude of the electric field strength $E_0$, which has been chosen, is:

$E_0 = 300$ kV / cm and 30 V / μ. At step h = 7 μ. There is a danger of interterm dielectric breakdown since the interterm voltage is 30 V / μ * 7 μ = 210 V. The breakdown voltage of the polyimide is 300 MV / m [4], page 550, or 300 V / μ, so that we can choose the helix structure as follows: the copper coil with a cross section Δ = 6 μ and polyimide insulation thickness of 1 micron.

*1. 4. 7. Electric power damping capacity while pulse propagating in the spiral*

The wave attenuation in the helical waveguide will lead to the fact that the amplitude of the pulse propagating in the spiral will decrease while pulse propagating from the beginning to the end of the spiral. This decrease is related with the Ohm current losses for heating the spiral.

$I_\varphi$ current flows through the spiral up and, in fact, the Ohm losses are as follows:

$$\Delta P \text{ (W/turn)} = \tfrac{1}{2} I_\varphi^2 * R, \qquad (25)$$

where: $I_\varphi$ - current expressed in amperes, R is the loop resistance in Ohms. Then Δ P / coil - is expressed in Watts.

First, we find the electrical resistance of one coil. Resistance is calculated according to the conventional formula: R = ρl / s, where ρ = 1.7 * $10^{-6}$ Ohm * cm, the resistivity of copper: l = $2\pi r_0$ - coil length, $r_0$ - radius of the helix, s - cross-section of the coil. Since the current flowing through the spiral is alternative (AC), then in formula (25) factor ½ appears. The alternative current in the conductor penetrates to the depth of the skin - layer, which must be found.

The expression for this depth of the skin - the layer can be written as follows:

$$\delta = c/(\sqrt{2\pi\sigma\omega_0}), \qquad (26)$$

where: c = 3 x $10^{10}$ cm / s - the velocity of light in vacuum , σ = 5.4 * $10^{17}$ 1 / s - conductivity of copper , $\omega_0 = 2\pi f_0$ - circular frequency , $f_0 = 43$ kHz - frequency of the wave propagating in the spiral. Substitution of the numerical values in formula (26) gives: δ = 0.03 cm.

The obtained skin depth - layer δ = 0.03 cm, much larger than the distance between the coils of the spiral h = 7 * $10^{-4}$ cm. This means that to reduce the resistance of one and, accordingly, to reduce attenuation of the coil, it is necessary to wind a rather wide tape with width $H = 2\delta = 0.06$ cm. Recall that we have chosen the ribbon thickness equal to $\Delta = 6$ μ.

Then the resistance of one coil $R = \rho l / s$ will be equal to:

$$R = \rho * 2\pi r_0/(2\delta * \Delta) = \rho * \pi r_0/(\delta * \Delta). \tag{27}$$

Substituting the numerical values for the start of the spiral, we find:

$$R = 1.7*10^{-6}*3.14/(3*6*10^{-6}) = 0.3 \text{ Ohm}. \tag{28}$$

Now we find $I_\varphi$ - current flowing through the coils. To do this, we use the following formula:

$$H_{zsurf} = (4\pi/c)nI_\varphi, \tag{29}$$

where: $H_{zsurf}$ - the magnetic field on the surface of the helix.

We find the relation between the component of the electric field $E_0$ on the helix axis and the component of the magnetic field $H_{z\,surf}$ on the spiral surface: $H_{z\,surf} = (k_1 / k) tg\Psi * I_0 (k_1 r_0) E_0 I_0 (k_1 r) / I_1 (k_1 r_0)$, [2]. For the inner area of the spiral, where $k_1$ - transverse wave vector: $k = (\omega / c) * \varepsilon^{1/2}$ - the wave vector, $r_0$ - the radius of the helix, the expression $(k_1 / k)$ is equal to: $(k_1 / k) = 1/\beta_{ph}$, $tg\Psi \approx h/2\pi r_0$, so that $(k_1 / k) * tg\Psi = \varepsilon^{1/2}$. For $k_1 r_0 = 1$ the ratio $I_0 (k_1 r_0) / I_1 (k_1 r_0) = 2$, i.e. $H_{z\,surf} = 2\varepsilon^{1/2} E_0$.

Thus, the component of the electric field: $E_0 = 300$ kV / cm, on the helix axis corresponds to the magnetic field strength $H_{z\,surf} = 70$ kGs on the surface of the helix.

Now we can find the current flowing through the coils of the spiral. The $nI_\varphi$ current can be found from the relation: $nI_\varphi$ (A / cm) $= H_{zsurf} / (4\pi / c) =$ $= (1.226)^{-1} * H_{zsurf}$ (A / cm) $= H_{zsurf}$ (Gs). Thus, the current in one coil is equal to:

$$I_\varphi (A) = H_{zsurf}(Gs)/n, \tag{30}$$

where: $n = 1 / h$ - the number of coils per one cm of the spiral.

In our case, n = 1/7 μ = 1.43 * 10³ coils / cm.

Substituting the numerical values in the formula (30), we find that the current in one coil is: $I_\varphi$ (A) = [70 kA / cm] / (1.43 * 10³ turns / cm) ≈ 50 A / turn. The Ohm losses of the current in one coil are equal to the following:

$$\Delta P \text{ (W/turn)} = \tfrac{1}{2} I_\varphi^2 * R = 375 \text{ W/coil}. \tag{31}$$

Since there are n coils per 1 cm, the energy losses per 1 cm will be by n times more than in one coil:

$$\Delta P \text{ (W/cm)} = \tfrac{1}{2} I_\varphi^2 * R * n = 536 \text{ kW/cm}. \tag{32}$$

We introduce the ratio:

$$\Delta P / P = -2\alpha, \tag{33}$$

whence:
$$1/\alpha = L_{damping} = 2P/\Delta P = 2*12.6*10^6 / 0.536*10^6 = 47 \text{ cm}. \tag{34}$$

This is the length over which the electric field strength is reduced by factor e due to attenuation. It can be seen that the motion of the balls when accelerating should be calculated taking into account the electric power. The accelerator itself should be divided into separate sections.

*1. 4. 8. Radial motion of the balls*

When particles are accelerated in an azimuthally symmetric field, it is not possible to achieve simultaneously both the radial and phase stabilities in the wave field [5]. The region of the phase stability corresponds to radial defocusing. The radial motion has the following form:

$$r(t) = r_{in} \exp[\gamma t], \tag{35}$$

i.e., any initial deviation from the acceleration axis $r_{in}$ exponentially increases with time. The growth increment γ is equal to [5]:

$$\gamma = \pi f_0 \{W_\lambda \text{ctg } \varphi_s / \pi \beta_{ph}\}^{1/2}, \tag{36}$$

where: $W_\lambda = (Z/A) eE_0\lambda_0 \sin\varphi_s / Mc^2$ is the energy rate at a wavelength in vacuum. We substitute the numbers for the beginning of the acceleration and find:

$$W_\lambda = 10^{-7}*3*10^5 *7*10^5*0.87/10^9 = 1.8*10^{-5},$$

$$\gamma = 3.14*4.28*10^4 \{1.8*10^{-5}*1.74/3.14*4.5*10^{-6}\}^{1/2} = 2*10^5.$$

The inverse value of $\gamma$ corresponds to the rise time of the radial deflection by factor e, $1/\gamma = 5*10^{-6}$ s. If we recall that the initial velocity of the iron ball is $V_{in} = 1.35$ km / s, it is possible to find the spatial increment of the initial deviation from the following relation:

$$l_{defl} = (1/\gamma)* V_{in} = 0.675 \text{ cm}. \qquad (37)$$

We get the length of the initial rise of the spatial deviation to be much smaller than the acceleration length $L_{acc} = 2$ m, that means that in this case it is necessary to introduce radial focusing. Let us consider focusing of the accelerated iron balls by means of quadrupole lenses.

It is known [5], the quadrupole lens focusing the particles in one plane defocuses them in another one. If two lenses are turned relatively each other by $90^0$, then in each of the transverse planes there are areas of the accelerator which alternately generate defocusing and focusing. Under certain conditions, such a system of lenses turns out to be the focusing.

Indeed, a particle moving accurately along the axis is not affected by any forces. The farther from the axis is the particle, the bigger values are the effective forces. Let the particle hit the first focusing section. Then it will bend the trajectory so to pass the defocusing section at a lower value of the field and the focusing forces will be stronger than the defocusing ones. A similar effect occurs in the case when the particle is first passing via the defocusing site. The resulting effect of the pair of the quadrupole lenses will be collecting, [5].

Focusing and defocusing action of the lenses is determined by their rigidity:

$$K = [(Z/A)eGl_l^2/Mc^2\beta_z], \qquad (38)$$

where $(Z/A)$ is the ratio of the charge to the mass, G - gradient of the electric or magnetic field in the lens, $l_l$ – lens length, $Mc^2 = 1$ GeV - the rest mass of the nucleon, $\beta_z$ - expressed in units of the velocity of light, the longitudinal velocity of the ball. Rigidity in this setting is unitless.

In contrast to the pair of quadrupole lenses, the accelerating section works as lens defocusing in the both perpendicular planes. Near the axis of acceleration there is no electric space charge and the following the condition:

$$\text{div } E = 0, \qquad (39)$$

it shows the ratio between the longitudinal and transverse electric fields:

$$E_r = - (r/2)dE_z/dz, \qquad (40)$$

This is evident, however, from the structure of the field in the spiral waveguide [2].

By analogy with the quadrupole lenses for the accelerating field the gradient is: $G_s = \frac{1}{2} dE_z / dz = \pi E_0/\lambda_{slow}$, where $E_0$ - the amplitude of the accelerating field, $\lambda_{slow}$ - slow wavelength in the structure.

At first we consider focusing of particles by electrostatic quadrupole lenses. We require that the rigidity of the quadrupole lenses to have a greater rigidity of the accelerating section. This means that the particles in the deflection angle lens must be larger than in the deflection angle of the section. Indeed, the angle of deflection of the particle in the accelerating section is always directed outside the section. The section defocuses the accelerated particles in both transverse planes. The particle on the focusing plane article focusing on the plot should get the deflection angle inside to be at the same location on the focusing plane, she will receive an additional deflection angle outward. [5]

This means that the gradient field in the lens multiplied by the square of its length must be greater than the gradient field in the accelerating section, also multiplied by the square of its length:

$$G_l * l_l^2 > G_s * l_s^2. \qquad (41)$$

The most difficult conditions for focusing occur at the beginning of the accelerator.

Let the lens length to be one-third of the length of the section, i.e. $l_s^2/l_l^2 = 10$, $l_s = 3l_l$. Then, the electric field gradient in the electrostatic lenses must at least be 10 times bigger than the electric field gradient in the sections, i.e:

$$G_l > 10\, G_s. \tag{42}$$

Substituting the numbers to start the acceleration where $\lambda_{slow} = 3.14$ cm, we find that the electric field gradient in the electrostatic lenses must be greater than: $G_l > 3 * 10^6$ V/cm$^2$.

The ratio between the focusing by the electrostatic quadrupole lenses and magnetic quadrupole lenses can be represented as follows [5]:

$$G_e\,(V/cm^2) = 300\, \beta\, G_m(Gs/cm), \tag{43}$$

where this coefficient 300 corresponds to the transition from units in Gs units to the units in V / cm. Coefficient $\beta$ is expressed in terms of the velocity of light, the longitudinal velocity of the particle. It is included into this formula as well as it enters the Lorentz force.

Substituting the value of $\beta = 4.5 * 10^{-6}$, we find that the magnetic field gradient in the quadrupole lenses must satisfy the following condition: $G_m > 2.2 * 10^9$ Gs / cm.

Let the length of the accelerating section be equal to: $l_s = 3$ cm. This length should be smaller than the attenuation length $L_{damping} = 47$ cm, which in this case is fulfilled. Then the length of the quadrupole lens is $l_l = 0.3\, l_s = 1$ cm. Now we can calculate the stiffness of the lens $G_l * l_l^2$, expressed in more familiar units –i.e. Tesla multiplied by meter. In this case, the value is as follows: $G_l * l_l^2 = 22\,T * m$.

Such high gradients in electrostatic and magnetic quadrupole lenses are "payment" for the high rate of acceleration. If such field gradients in the lenses are rather difficult to obtain, it is necessary to use a lower rate of acceleration. Placing of the focus elements increases the length of the accelerator by approximately twice.

*2. Acceleration of the beryllium sphere*

We have to say that from the point of view of accelerating the solid body acceleration is not optimal. This is due to the fact that for a solid body with the increasing radius of the ball a very important parameter Z / A. rapidly decreases. The charge on the bead is distributed on its surface, while the mass is concentrated in the ball volume and the ratio of the surface to the volume becomes smaller with increasing of the ball radius.

If from the acceleration of the solid ball to go to the acceleration of the sphere where all the mass is concentrated near the surface of the sphere, it can be expected that the ratio Z / A in this case is much larger than for the ball. We will consider the conditions of accelerating the beryllium sphere with radius $r_{Be}$ = 10 μ, the same as for the iron ball. The equal thickness of the sphere: $δ_{Be}$ = 1 μ. is taken.

*2 . 1. The number of protons placed on the sphere*

We find the number of protons located on the sphere on the Coulomb law:

$$E_{surf} = N_p \, e/r_{Be}^2,$$

where $E_{surf}$ = 300 MV / cm – the surface electric field strength on the sphere,
e = 4.8 * $10^{-10}$ – is the elementary charge, expressed in CGS units,
$r_{Be}$ = 10 μ - radius of the sphere, $N_p$ - the number of protons located on the sphere.

Substituting numbers into this formula, we find that to obtain the surface electric field strength $E_{surf}$ = 300 MV / cm, it is required to have the following number of protons: $N_p$ = 2 * $10^9$. It is clear that for the same surface electric field strength and the same radius of the sphere and of the ball we get the same number of protons, which should be placed on the sphere or the ball.

*2. 2. Number of nucleons contained in the sphere*

We find the mass of the beryllium sphere. The volume of the sphere is equal to: $V_{Be}$ = 4π $r^2_{Be}$ * $δ_{Be}$ ≈ 1.26 * $10^{-9}$ cm$^3$, the density of beryllium is assumed to be $ρ_{Be}$ = 1.84 g/cm$^3$, so that the mass of the ball is $m_{Be}$ ≈ 2.3 * $10^{-9}$ g. The number of nucleons in the sphere is found from Avogadro's law:

$$6*10^{23} \text{ ------------ } 9 \text{ g}$$

$$x \text{ ------------} 2.3*10^{-9} \text{ g,} \qquad (44)$$

where: 9 g - mass of one gram mole of beryllium. From (44) we find that
$x = 6 * 10^{23} * 2.3 * 10^{-9}/9$ atoms or $A = 6 * 10^{23} * 2.3 * 10^{-9} =$
$= 1.4 * 10^{15}$ nucleons.

The ratio of the charge to the mass (Z / A) for the sphere is equal to the following:

$$eN_p/A = Z/A = 2*10^9/1.4*10^{15} = 1.4*10^{-6}. \qquad (45)$$

So we have obtained a the ratio of the charge to the mass to be by 14 times greater than that for the iron ball, that means that for the same accelerator parameters it is 14 times shorter.

**3. Acceleration of the tungsten tube segment**

It is necessary to note that concerning the depth of penetration of substances into the body, the spherical shape of the body is not optimal. To increase the penetration depth, it is needed to use an elongated arrow-like -shaped body. Thus, it is possible to increase the mass of the body without increasing its cross-section.

Let us consider the conditions to accelerate a segment of the tungsten tube with a diameter $d_W = 20$ μ, wall thickness $\delta_W = 1$ μ and length $l_W = 1$ cm.

*3. 1. The number of protons placed on the segment of the tube*

We find the number of protons placed on the segment of the tube according to the Coulomb law:

$$E_{surf} = (2\kappa e/r_W), \qquad (46)$$

where $E_{surf} = 300$ MV / cm is the surface electric field strength on the segment of the tube, $e = 4.8 * 10^{-10}$ is the elementary charge, expressed in CGS units, $r_W = 10$ μ -the segment radius, κ - the number of protons per one centimeter length of the tube. Substituting the numbers into this formula in the CGS system:

$$10^6 = 2 \, e \, \kappa / 10^{-3},$$

we find that it is required to have the number of protons equal to: $\kappa = N_p = 10^{12}$ to reach the surface electric field strength $E_{surf} = 300$ MV/cm. The tube section should be covered on the front and back parts with the tungsten hemispheres to have no sharp edges of the segment of the tube where large electric overvoltage could be expected.

*3. 2. Number of nucleons in the tube segment*

We find a mass of the tungsten tube segment. The volume of the segment is equal to: $V_W = 2\pi \, r_W * \delta_W * l_W \approx 6.28 * 10^{-7}$ cm$^3$, the density of tungsten is assumed to be $\rho_W = 19$ g/cm$^3$, so that the mass of the segment is: $m_W \approx 1.2 * 10^{-5}$ g. The number of nucleons in the tube segment is found from Avogadro's law:

$$6*10^{23} \text{ ------------ } 184 \text{ g}$$

$$x \text{ ------------ } 1.2*10^{-5} \text{ g,} \qquad (47)$$

where 184 g - mass of one gram mole of tungsten. From (47) we find that
$x = 6 * 10^{23} * 1.2 * 10^{-5}/184$ atoms or $A = 6 * 10^{23} * 1.2 * 10^{-5} = 7.2 * 10^{18}$ nucleons.

The ratio of the charge to the mass of the nucleon is $(Z / A)$, thus the tube segment is equal to:

$$eN_p/A = Z/A = 10^{12}/7.2*10^{18} = 1.4*10^{-7}. \qquad (48)$$

The obtained ratio of the charge to the mass is 1.4 times greater than for the iron ball that means that for the same parameters of the accelerator it will be by 1.4 times shorter than the accelerator for the iron balls, namely: $L_{acc \, W} = 1.4$ m.

**4. Interaction of artificial micrometeorites with the magnetic and gravitational fields of the Earth**

When moving in the Earth magnetic field the charged iron ball, like any charged particle, will be deviated by this magnetic field.
We take the magnetic path length approximately equal to the radius of the Earth $L_m = 7 * 10^3$ km, the magnetic field of the Earth is assumed to be $H_E = 0.5$ Gs.

Then the deflection angle in the magnetic field and for this length of the iron ball will be equal to:

$$\Theta_m = (Z/A)e\, L_m * H_E/Mc^2 = 10^{-7}*7*10^8*0.5*3*10^2/10^9 \approx 10^{-5}. \quad (49)$$

It is clear that the magnetic field of the Earth does not influence the motion of the iron ball.

For the same length of the track $L_{grav} = L_m = 7 * 10^3$ km, we find the deflection angle of the iron ball by the gravitational field of the Earth is also very small:

$$\Theta_{grav} = g\, L_{grav}/V^2_m = 10^{-2}*7*10^3/30*30 = 7.7*10^{-2}. \quad (50)$$

**5. The depth of penetration of artificial micrometeorites in the Earth atmosphere**

We assume that the main reason of destruction of artificial micrometeorite in the Earth atmosphere is their heating and evaporation due to this heating. It is assumed that the same amount of energy is needed for heating micrometeorite and atmospheric gas of the Earth.

*5. 1. Penetration of the iron ball into the Earth atmosphere*

We calculate the motion of the ball, taking into account the air resistance. The equation of the motion of the ball can be written in the following form (excluding the force of gravitational attraction to the Earth):

$$m dV/dt = -\rho C_x S_{tr} V^2/2, \quad (51)$$

where: m – is the mass of the ball, V-velocity, $\rho = \rho_0 e^{-z/H0}$ - barometric formula changes in atmospheric density with altitude $\rho_0 = 1.3 * 10^{-3}$ g/cm$^3$ - the air density on the Earth surface, $H_0 = 7$ km - the height at which the density decreases by factor e, $C_x$ - drag coefficient. Let us assume that the value for the ball is: $C_x = 1$. In our case for the ball, $S_{tr} = \pi r^2_{Fe} = 3.14 * 10^{-6}$ cm$^2$. The solution of equation (51) can be written as follows:

$$V(t) = V_m/[1+ \rho V_m * S_{tr} * t/2m]. \quad (52)$$

If to sum up the energy required for heating the ball from the room temperature to the melting point [4], page 289, the energy of the phase transition "solid – liquid", the energy that must be used for heating of the ball till its boiling point and, finally, the energy of the phase transition of the liquid to vapor [4], page 289, we will get the value of energy: $\Delta E_{Fe} = 7.9$ kJ / g.

The kinetic energy of the ball can be found from the following considerations: the energy per nucleon in the ball is 5 eV / nucleon, the ball contains $2 * 10^{16}$ nucleons, so that the total energy is equal to $10^{17}$ eV or 0.624 J.

Multiplying the energy that must be expended to evaporate one gram of iron per the weight of the iron ball $m_{Fe} = 3 * 10^{-8}$ g, we find that for the evaporation of the iron ball it is needed to spend:
$\Delta W_{evap} = 7.9$ kJ / g $ * 3 * 10^{-8}$ g $= 2.4 * 10^{-4}$ J. This is a very small part of the kinetic energy of the ball: $W_{kin} = 0.624$ J.

The energy losses are related with the losses of the velocity by the following ratio: $\Delta E / E = 2 \Delta V / V$. It means that when the relative velocity decreases $\Delta V / V$ the amount $\Delta W_{evap} / W_{kin} = 2.4 * 10^{-4}/0.624 = 3.8 * 10^{-4}$, the iron balls evaporate. The magnitude of $\Delta V / V$ can be determined from the formula (52):

$$\Delta V/V = \rho V_m * S_{tr} * t / 2m. \tag{53}$$

After replacing $V_m dt$ by $dz$, we will integrate the value $\infty$ to $z_0$, where $z_0$ is the height, measured from the surface of the Earth, where the iron ball will evaporate. When integrating we obtain the following:

$$H_0 * \int_\infty^{(z_0/H_0)} \exp(-z/H_0) \, d(z/H_0) = H_0 \exp(-z_0/H_0). \tag{54}$$

Thus, we have obtained an exponential equation to determine the value of $z_0$:

$$\rho_0 * S_{tr} * H_0 \exp(-z_0/H_0) / 2m = \Delta W_{evap} / W_{kin}, \tag{55}$$

where: $\rho_0 = 1.3 * 10^{-3}$ g/cm$^3$, $S_{tr} = \pi r^2_{Fe} = 3.14 * 10^{-6}$ cm$^2$, $H_0 = 7$ km.

Substituting numbers into the equation (55) we obtain:

$$1.3 * 10^{-3} * 3.14 * 10^{-6} * 7 * 10^5 * \exp(-z_0/H_0) / (2 * 3 * 10^{-8}) = 3.8 * 10^{-4},$$

From where we find:

$$\exp(-z_0/H_0) = 3.8 \cdot 10^{-4}/5.15 \cdot 10^4 \approx 10^{-8},$$

$$0.434 \cdot (-z_0/H_0) = -8, \quad z_0 = 130 \text{ km}.$$

Thus, at the height $z_{0\,Fe} = 130$ km from the Earth surface the iron ball evaporates.

*5. 2. Penetration of the beryllium sphere into the Earth atmosphere*

Summing up all the energy in a similar way as for beryllium, we find that the energy that must be expended to vaporize one gram of beryllium is equal to the following: $\Delta E_{Be} = 38$ kJ / g [5], page 289.

Taking into account the same reasons we find the kinetic energy of the sphere. The energy per nucleon in the sphere is: 5 eV / nucleon, the sphere contains $1.4 \cdot 10^{15}$ nucleons, so that its kinetic energy is equal to $7 \cdot 10^{15}$ eV or $W_{kin} = 4.37 \cdot 10^{-2}$ J.

Multiplying the energy that must be expended to evaporate one gram of the beryllium sphere by the mass of the beryllium sphere $m_{Be} = 2.3 \cdot 10^{-9}$ g, we find that for evaporation of the beryllium sphere the following energy must be expended: $\Delta W_{evap} = 38$ kJ / g $\cdot 2.3 \cdot 10^{-9}$ g $= 8.7 \cdot 10^{-5}$ J. Substituting the numbers into equation (55), we find that the ratio $\Delta W_{evap} / W_{kin} = 2 \cdot 10^{-3}$,

$$1.3 \cdot 10^{-3} \cdot 3.14 \cdot 10^{-6} \cdot 7 \cdot 10^5 \cdot \exp(-z_0/H_0) /(2 \cdot 2.3 \cdot 10^{-9}) = 2 \cdot 10^{-3},$$

$$6.2 \cdot 10^5 \exp(-z_0/H_0) = 2 \cdot 10^{-3}, \ \exp(-z_0/H_0) = 3 \cdot 10^{-9}, \ 0.434 \cdot (z_0/H_0) \approx 9.$$

Where $z_0 = 150$ km, i.e., the beryllium sphere evaporates at height $z_{0\,Be} = 150$ km from the Earth surface.

*5. 3. Penetration of the tungsten tube segment into the Earth atmosphere*

We determine the kinetic energy of the tungsten tube segment and define the amount of energy in the electron – volts taking into account that the tube section contains $7.2 \cdot 10^{18}$ nucleons, and the kinetic energy per nucleon is 5eV. Finally, we obtain $E_{kin} = 7.2 \cdot 10^{18} \cdot 5 = 3.6 \cdot 10^{19}$ eV. Multiplying this value by $6.24 \cdot 10^{-18}$, we obtain the value of the kinetic energy in Joules: $W_{kin} = 224$ J. The value of the energy that must be expended to evaporate one gram of tungsten is: [4], page 289, $E_{evap} = 5$ kJ / g, to vaporize the tungsten tube segment with mass $m_W = 1.2 * 10^{-5}$ g, it is needed to spend $W_{evap} = 6 * 10^{-2}$ J. This means that $W_{evap} / W_{kin} = 2.7 * 10^{-4}$ of the kinetic energy of the tube segment must be expended for evaporation. Substituting this value in equation (55), we obtain:

$$1.3*10^{-3}*3.14*10^{-6}*7*10^{5}*\exp(-z_0/H_0)/(2*1.2*10^{-5}) = 2.7*10^{-4},$$

$$\exp(-z_0/H_0) = 10^{-6}, \quad 0.434*(z_0/H_0) = 6, \quad z_0 = 100 \text{ km}.$$

This means that the length of the tungsten tube evaporates at height $z_{0\,W} = 100$ km from the surface of the Earth.

*5.4. Penetration of the tungsten tube in aluminum*

We take the density of aluminum equal to $\rho_{Al} = 2.7$ g/cm$^3$ [4], page 99, the mass of one gram moles of aluminum is 27 g. The specific heat of aluminum is, $c_{Al} = 24.35$ J / (mol * K) [4], p.199, the melting point $T_{mel} = 660$ C$^0$, heat of phase transition solid - liquid is, $\Delta H_{mel} = 10.8$ kJ / mol, [4], page.289, the boiling point is $T_{eva} = 2520$ C$^0$. To reach the boiling point of aluminum, it is necessary to expend 50 kJ / mol, the heat for the phase transition "liquid – vapor" is $\Delta H_{eva} = 293$ kJ / mol, [4], p. 289.

If you sum up all the required energy, we find that the evaporation of one mole of aluminum requires 375 kJ; the evaporation of one cubic centimeter of aluminum is 37.5 kJ/cm$^3$. The cross-section of the tungsten tube segment is $S_{tr\,W} = 2\pi r_W * \delta_W = 6.28 * 10^{-7}$ cm$^2$, so, to evaporate one centimeter of aluminum, it is required to spend: $W_{evap\,1\,cm\,Al} = 2.35 * 10^{-2}$ J / cm. Recall that evaporation of the tungsten tube segment consumes $W_{evap\,W} = 6 * 10^{-2}$ J. Assuming that the kinetic energy of the tube segment is consumed in equal proportions for evaporation of tungsten and aluminum, we find the depth of penetration of the segment in the aluminum, after which the tungsten tube segment evaporates: $l_{W\,evap}$,

$$l_{W\,evap} = W_{W\,evap} / W_{evap\,1\,cm\,Al} = 6*10^{-2}/2.35*10^{-2} = 2.5 \text{ cm}. \quad (56)$$

It is seen that, in this case, the main reason limiting the depth of penetration of the body into the material is not the loss of kinetic energy of the body but its evaporation.

*5.5. Penetration of the beryllium segment into the Earth atmosphere*

We consider penetration into the atmosphere of the beryllium tube segment having the same dimensions as tungsten: the diameter of the tube $d_{Be} = 20$ μ, the thickness of the wall $\delta_{Be} = 1$ μ, the length $l_{Be} = 1$ cm. Recall that in front and at the end this segment of the tube should be covered by hemispheres to avoid sharp edges of the segment. The mass of such a segment is less than the mass of the tungsten segment in respect of specific weights of tungsten and beryllium and it is equal to: $m_{Be\ tube} = 1.15 * 10^{-6}$ g. The segment contains: $A_{Be\ tube} = 7 * 10^{17}$ nucleons.

We assume that the beryllium tube segment is electrically charged till the same surface electric field strength - $E_{surf} = 300$ MV / cm, for this purpose we have to place $N_p = 10^{12}$ protons. The ratio of the charge to the mass for this segment will be equal to $Z/A = 1.43 * 10^{-6}$.

For the beryllium tube segment the ratio of the charge to the mass is approximately the same as for the beryllium sphere.

The kinetic energy of the segment is: $E_{kin} = 7 * 10^{17} * 5 = 3.5 * 10^{18}$ eV. Transferring this value in Joules we obtain $W_{kin} = 3.5 * 6.24 = 22$ J.

To vaporize the beryllium tube segment, it is needed to consume $\Delta W_{evap} = 38$ kJ / g $* 1.15 * 10^{-6}$ g $= 4.37 * 10^{-2}$ J. The ratio of the energy required for evaporation to the kinetic energy of the tube segment is as follows:

$$\Delta W_{evap} / W_{kin} = 2*10^{-3}.$$

Substituting this value in the right-hand side of equation (55), we obtain:

$$1.3*10^{-3}*3.14*10^{-6}*7*10^{5}*\exp(-z_0/H_0)/(2*1.15*10^{-6}) = 2*10^{-3},$$

$$\exp(-z_0/H_0) = 10^{-6},\ 0.434*(z_0/H_0) = 6,\ z_0 = 100 \text{ km}.$$

The beryllium tube segment will evaporate at height $z_{0\ Be\ tube} = 100$ km, while the beryllium sphere evaporates at height $z_{0\ Be} = 150$ km, that shows the advantage (from the point of view of penetration into the material) of the elongated shape in comparison with the spherical body.

### 5.6. Penetration of the beryllium tube segment with a graphite cone in the head part

In [6] it is shown that using a sharp cone in the cylinder head results in significant reduction in the drag coefficient $C_x$. The connection between the cone angle, $\Theta$ head, and drag coefficient is given by [6]:

$$C_x = \Theta^2_{head}. \qquad (57)$$

Thus, for the values of the angle $\Theta_{head} = 10^{-3}$ the obtained drag coefficient $C_x = 10^{-6}$.

Into the head of the beryllium tube segment which is electrically insulated from it, we place a graphite cone with length $l_{cone} = 2$ cm. Let the diameter of the cone $d_C$ coincide with the diameter of the beryllium tube segment $d_C = 20$ μ, and let it be equal to the thickness of $\delta_C = 1$ μ. The cone of this length is the very cone: $\Theta_{head} = d_C / l_{cone} = 10^{-3}$.

The lateral surface area of the cone is: $S_{cone} = (½) \pi d_C l_{cone} = 6.28 * 10^{-3}$ cm$^2$, volume $V_{cone} = 6.28 * 10^{-7}$ cm$^3$, mass $m_{cone} = 1.4 * 10^{-6}$ g, where we took the density of graphite to be equal to $\rho_{grap} = 2.25$ g/cm$^3$, [4], page 100. This cone contains $A_C = 6 * 10^{23} * 1.4 * 10^{-6} = 8.4 * 10^{17}$ nucleons. Then the total number of nucleons in the beryllium tube segment and graphite cone will be equal to: $A_{total} = (7 + 8.4) * 10^{17} = 1.54 * 10^{18}$ nucleons and the ratio of the charge to the mass is equal to $Z / A = 10^{12}/1.54 * 10^{18} = 6.5 * 10^{-7}$, that is approximately by 6.5 times greater than for the iron ball. This means that the accelerator for the segment of the beryllium tube with a sharp graphite cone at the head part of this segment is by 6.5 times shorter than the accelerator for the iron bead, whose length is equal to: $L_{acc\ Fe} = 2$ m.

Taking the specific heat of graphite $C_C = 8.54$ J / (mol * degree) [4], page 199 and the temperature at which graphite is destroyed, to be equal to: $T_{des\ C} = 3700$ C$^0$, we find that for the destruction of one gram of graphite it is needed to spend $W_{des\ C} = 8.54 * 3700/12 = 15.8$ kJ / g. To destroy the cone, it will be necessary to spend $W_{C\ cone} = W_{des\ C} * m_{cone} = 2.2 * 10^{-2}$ J.

The total energy of the system plus the beryllium tube segment with a cone is as follows: $A_{total} * 5$ eV $= 1.54 * 10^{18} * 5$ eV $= 7.7 * 10^{18}$ eV or $W_{kin} = 7.7 * 6.24 = 48$ J. Thus, the ratio of the energy required to destroy the cone, to the total kinetic energy of the system is equal to:

$W_{C\ cone} / W_{kin} = 2.2 * 10^{-2}/48 = 4.6 * 10^{-4}$. The total mass of the system is: $m_{total} = 2.55 * 10^{-6}$ g.

Now we substitute all the numbers in the equation (55), where additional factor $C_x = 10^{-6}$ appears:

$$10^{-6}*1.3*10^{-3}*3.14*10^{-6}*7*10^5*\exp(-z_0/H_0)/(2*2.55*10^{-6}) = 4.6*10^{-4},$$

$$\exp(-z_0/H_0) \approx 1, \quad z_{0\ comp} = 0.$$

This means that the above composite micrometeorite (beryllium tube segment plus a graphite cone) will reach the Earth surface without being destroyed.

**Conclusion**

Artificial micrometeorites can be a convenient tool to study the Earth atmosphere.

Literature

1. Tables of physical quantities, Handbook ed. I. K. Kikoin, Moscow, Atomizdat, 1976

2. S.N. Dolya, KA Reshetnikova, On electrodynamics acceleration of macroscopic particles, Preprint, P9-2009-110, Dubna, 2009
http://www1.jinr.ru/Preprints/2009/110(P9-2009-110).pdf
http://arxiv.org/ftp/arxiv/papers/0908/0908.0795.pdf

3. S. N. Dolya, K. A. Reshetnikova, Linac ions $C^{+6}$ - injector synchrotron designed for hadrons therapy, Preprint JINR, P9-2011-82,
http://www1.jinr.ru/Preprints/2011/082(P9-2011-82).pdf
http://arxiv.org/ftp/arxiv/papers/1307/1307.6302.pdf

4. Physical quantities, Directory ed. I. S. Grigor'ev
and E. Z. Mey'likhov, Moscow, Energoatomizdat, 1991

5. I. M. Kapchinsky, Particle dynamics in linear resonance accelerators, Moscow, Atomizdat, 1966